\documentclass[conference]{IEEEtran}

\usepackage[utf8]{inputenc}
\usepackage{cite}
\usepackage{xcolor}
\usepackage{color,soul}
\usepackage{comment}
\usepackage{amsmath,amssymb,amsfonts}
\usepackage{algorithmic}
\usepackage{graphicx}
\usepackage{textcomp}
\usepackage{tikz}

\usepackage{pgf-pie}  
\usepackage{pgfplots} 
\pgfplotsset{compat=1.8}  
\usepackage{tikz}
\usepackage{hyperref}

\usepackage{enumitem} 
\usepackage{algorithm}

\usepackage{subfigure}
\usepackage[margin=1in]{geometry}  

\ifCLASSINFOpdf
\else
\fi
\begin{document}
%

\title{Multi-MedChain: Multi-Party Multi-Blockchain Medical Supply Chain Management System}

\author{
    \IEEEauthorblockN{Akanksha Saini\IEEEauthorrefmark{1}, Arash Shaghaghi\IEEEauthorrefmark{2}, Zhibo Huang\IEEEauthorrefmark{2}, Salil S. Kanhere\IEEEauthorrefmark{2}}
    \IEEEauthorblockA{
        \IEEEauthorrefmark{1}RMIT University, Melbourne, Australia\\
        \IEEEauthorrefmark{2}The University of New South Wales (UNSW), Sydney, Australia\\
        Email: akanksha.saini@rmit.edu.au
    }
}



%


 \maketitle

\begin{abstract}

The challenges of healthcare supply chain management systems during the COVID-19 pandemic highlighted the need for an innovative and robust medical supply chain. The healthcare supply chain involves various stakeholders who must share information securely and actively. Regulatory and compliance reporting is also another crucial requirement for perishable products (e.g., pharmaceuticals) within a medical supply chain management system. Here, we propose Multi-MedChain as a three-layer multi-party, multi-blockchain (MPMB) framework utilizing smart contracts as a practical solution to address challenges in existing medical supply chain management systems. Multi-MedChain is a scalable supply chain management system for the healthcare domain that addresses end-to-end traceability, transparency, and collaborative access control to restrict access to private data. We have implemented our proposed system and report on our evaluation to highlight the practicality of the solution. The proposed solution is made publicly available.

\end{abstract}



%
\IEEEpeerreviewmaketitle

\section{Introduction}
\begin{figure}[tbh]
\centering{}\textsf{\includegraphics[width=1\columnwidth,scale=10]{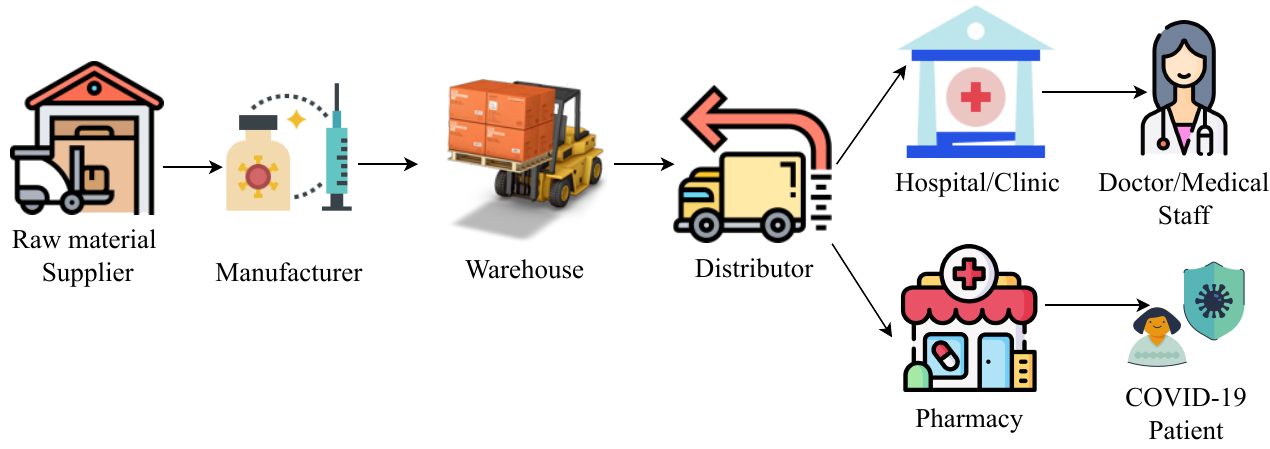}}\caption{The workflow of medical supply chain during COVID-19.} \label{fig: Supply chain.}
\label{fig: Current supply chain flowchart for COVID-19 medical supplies.}
\end{figure}

The healthcare supply chain comprises interconnected systems, components, and processes that collaboratively function to manufacture, distribute, and deliver medications and other healthcare supplies. A typical supply chain includes several organisations, resources, and interdependent operations. Fig. \ref{fig: Supply chain.} represents the workflow of the medical supply chain during COVID-19. With the outbreak of the COVID-19 pandemic, there was a sudden surge in demand for drugs, masks, and PPE kits, overwhelming the healthcare supply chain. The scarcity of personal protective equipment (PPE) and masks during the initial stages of the pandemic is a prime example of the consequences of an inadequate healthcare supply chain management system (SCM) \cite{park2020global}. The holding and black marketing of lifesaving drugs and equipment by some distributors for higher profits have further induced the Bullwhip effect in the supply chain, leading to mismatches between demand and production and decreased efficiency of the supply chain \cite{bray2012information}.



The efficient delivery of medical equipment and supplies, particularly in times of crisis, depends on effective supply chain management. The medical supply supply chain is managed by the healthcare supply chain management system (SCM) from the point of manufacture to the patient. SCMs are recognized to be hindered by the lack of information exchange between partners, which is ascribed to the presence of multiple supply chains with changing trading partners over time, reluctance to share competitive and private information, and the absence of direct contact \cite{nakasumi2017information}. Managing the supply chain encompasses a multitude of stakeholders, each with own goals and concerns, rendering it a multifaceted and intricate procedure. Consequently, a more intelligent, decentralized, and efficient supply chain is needed by the healthcare sector \cite{dutta2020blockchain}.

Poor product consistency, limited data reporting, a lack of automation, and rising regulatory constraints make healthcare SCM even more complicated. These issues raise supply costs that hospitals and providers struggle to manage.

There has been a growing interest among industry leaders and managers in adopting blockchain technology in healthcare supply chains \cite{pournader2020blockchain}.
With the rapid development of blockchain technology in healthcare, several blockchain-based mechanisms have been proposed to ensure traceability from one end to another across the supply chain \cite{kshetri20181}. However, the question of sharing accurate information and maintaining data privacy to ensure competitiveness in the market is a research challenge that has not been adequately addressed.
In this paper, we propose Multi-MedChain as a novel and robust three-layer multi-party, multi-blockchain (MPMB) healthcare supply chain management system. Our proposed solution enables secure cross-chain communication to establish digital trust while keeping sensitive information private on local blockchains. For instance, trading secrets of supply chain stakeholders, such as capacities and prices, must be kept private in such a system. To further limit and impose control over the sharing of private information for different parties involved in the SCM, we have devised a dynamic access control mechanism that leverages smart contracts. This allows supply chain stakeholders to collaboratively manage user authorisations and achieve end-to-end traceability. We have implemented our solution, and the evaluation results prove the scalability of Multi-MedChain. The project source code has been made publicly accessible for replication and extensions.

\section{Literature Review}

Medical and pharmaceutical companies are actively exploring blockchain-enabled solutions to provide a secure and reliable method of tracking their products throughout the supply chain. Modum \cite{cworld} operates on the Ethereum blockchain, powered by smart contracts, and checks the state of the drug at strategic stages to ensure that it meets the required standard. 
Chronicled \cite{dworld}, Blockpharma \cite{eworld} are some of the ongoing projects that use blockchain to redefine the medical supply chain.  


\begin{figure}[tbh]
\centering{}\textsf{\includegraphics[width=0.9\columnwidth,scale=25]{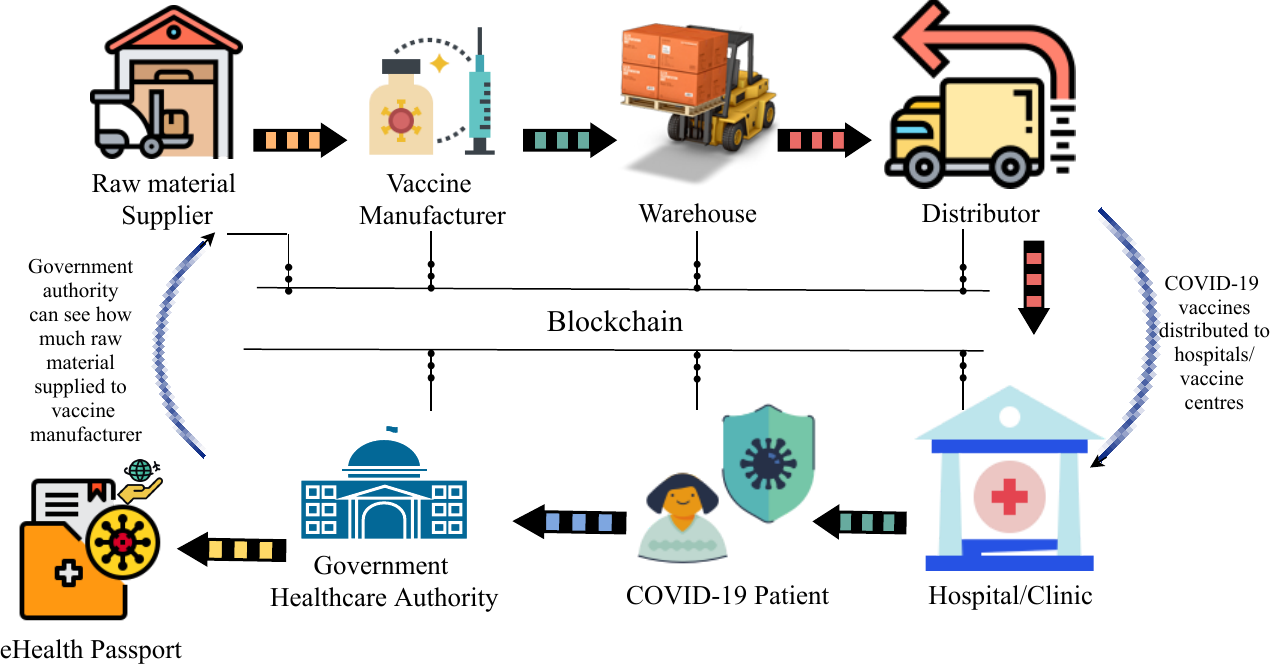}}\caption{Circular blockchain-based healthcare SCM.} \label{fig: circular supply chain}
\end{figure}

Blockchain technology in healthcare promotes efficiency and provides better control over supply chains. It also improves overall patient safety, solves the problem of drug authenticity and traceability, and enables secure interoperability among healthcare institutions \cite{reda2020blockchain}. In \cite{jamil2019novel} a blockchain-based drug SCM has been proposed using Hyperledger fabric to enhance the functionality of the supply chain and ensure the management of drug records. Similarly, blockchain can create a transparent and traceable vaccine supply chain, building trust between institutions and patients.  \cite{yong2020intelligent}. In \cite{alam2021challenges}, the authors list crucial challenges facing the COVID-19 vaccine supply chain, such as a) accurate prediction of vaccine demand, b) proper planning and scheduling, c) coordination with local organisations, and d) correspondence among SCM vaccine entities. A smart contract-based solution for safe vaccine distribution and tracking has been proposed in \cite{li2023novel}. In \cite{peng2020efficient}, a double-level blockchain structure has been designed to supervise vaccine production. Furthermore, a smart contract-based drug traceability mechanism has been proposed in \cite{musamih2021blockchain}.  \cite{sharma2023blockchain} has devised a distributed privacy-preserving application to maintain medical certificates.

As illustrated in Fig.~\ref{fig: circular supply chain}, all currently in use healthcare SCM systems operate in a circular fashion, with many parties sharing the same blockchain network. In these kinds of systems, one or more network participants control the consensus. Nevertheless, there are a number of possible disadvantages to this strategy, such as disagreements amongst participants, exposed proprietary information or intellectual property (IP) (demands, capacities, orders, pricing, margins), and a bias in favor of privacy in open access \cite{yang2021behavioural}. Conflicts often arise over the selection of the platform and information to be delivered.

The aforementioned works have not examined systems with several blockchains for each stakeholder. Single blockchain models in the healthcare supply chain ecosystem prevent stakeholders from sharing information directly, securely, and transparently in real time. Our framework supports cross-chain solutions amongst several blockchains to bypass existing framework constraints. Our technology, Multi-MedChain, helps the SCM build digital trust and protect client interactions. While helping multiple enterprises and governments create digital trust, these solutions give more control by directly accessing sensitive information for superior, more tailored services with enhanced security and privacy. Each party runs own blockchain network, eliminating consensus methods, membership, and technological stack disputes. 

The comprehensive comparison of related work in the blockchain-based healthcare supply chain \cite{kshetri20181,xu2019systematic,monrat2019survey,gong2021multiple,collier2021zero,xu2020data,clauson2018leveraging} is summarised in Table I. The comparison is conducted in terms of various key features that are enumerated in the first column of the table.

\begin{table}
\caption{Summary of key features of the existing works in healthcare supply chain.}
\resizebox{\columnwidth}{!}{
\begin{tabular}{|l|l|l|l|l|l|l|l|}
\hline
\textbf{Features}                      & [8]       & [22]       & [23]       & [24]       & [25]       & [26]       & [27]        \\ \hline
Anonymity                              & X         & X          & \checkmark & \checkmark & X          & X          & X           \\ \hline
Access control                         & X         & X          & \checkmark & \checkmark & X          & \checkmark & \checkmark  \\ \hline
Authentication                         & X         & X          & X          & \checkmark & \checkmark & \checkmark & X           \\ \hline
Data privacy                           & X         & X          & X          & X          & \checkmark & X          & \checkmark  \\ \hline
Information asymmetry                  & X         & X          & X          & X          & X          & X          & X           \\ \hline
Information sharing                    & X         & X          & \checkmark & \checkmark & \checkmark & \checkmark & \checkmark  \\ \hline
Multi-chain                            & X         & X          & X          & \checkmark & X          & X          & X           \\ \hline
Multi-layer                            & \checkmark& \checkmark & X          & \checkmark & \checkmark & X          & X           \\ \hline
Traceability                           & X         & X          & X          & X          & \checkmark & \checkmark & \checkmark  \\ \hline
Governance                             & X         & \checkmark & X          & X          & \checkmark & X          & X           \\ \hline
Cross-chain/Interoperability           & X         & X          & X          & \checkmark & X          & X          & X           \\ \hline
Sustainability                         & \checkmark& \checkmark & X          & X          & X          & X          & X           \\ \hline
\end{tabular}
}
\label{Table1}
\end{table}



\section{Proposed Solution: Multi-MedChain}



\begin{figure}[t!]
\centering{}\textsf{\includegraphics[width=1\columnwidth,scale=0.8]{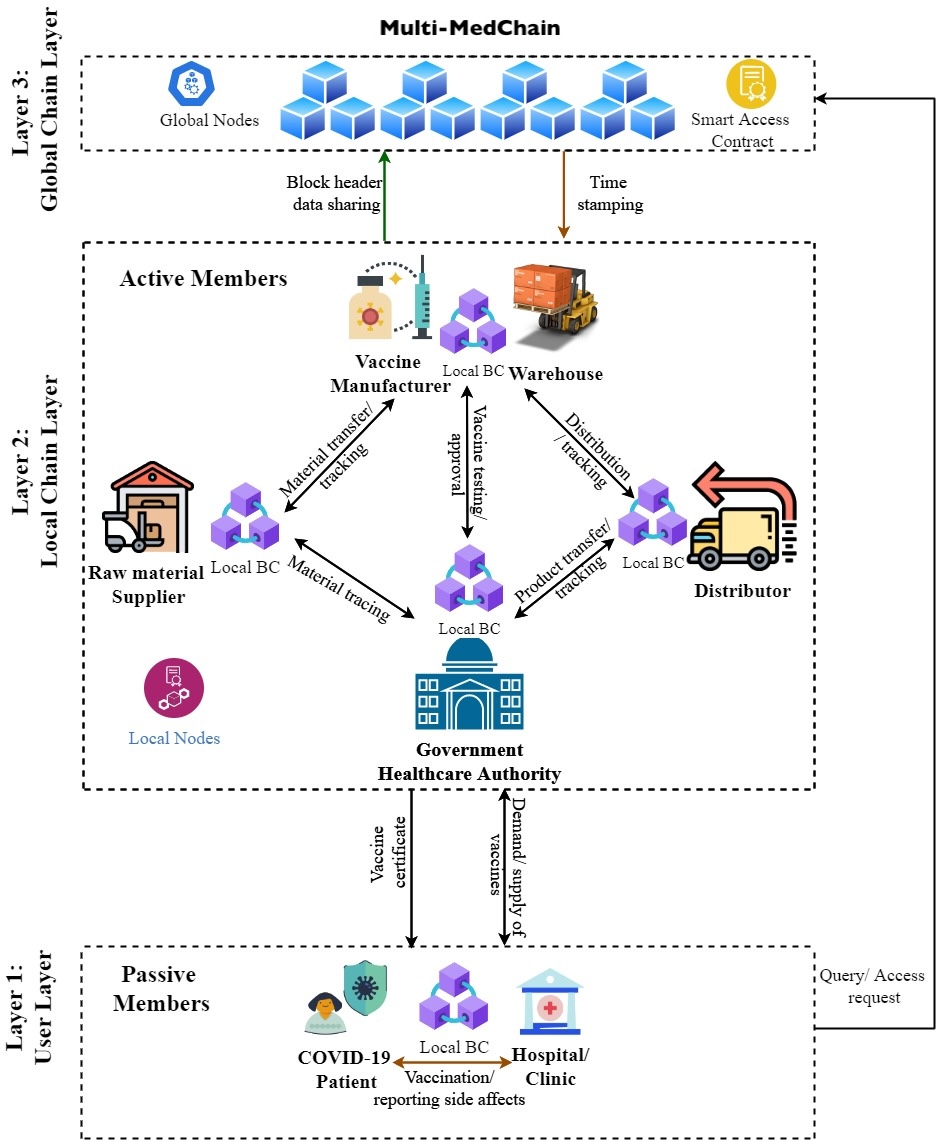}}\caption{Multi-MedChain framework.} \label{fig: proposed system}
\end{figure}

To facilitate demand-supply coordination, compliance certification, and improve information sharing in the healthcare supply chain management system, we propose an end-to-end solution using a multi-party multi-blockchain model illustrated in Fig.~\ref{fig: proposed system}. For simplicity, we focus on the vaccine supply chain and present Multi-MedChain. However, the proposed approach is adaptable to the supply chain of other medical supplies. 
 

\subsection{Entities and Components}

We describe the role of each stakeholder and component involved in each layer of our proposed system.

\begin{itemize}
   \item \textit{Passive Members.} Passive members mainly include patients, hospitals, and clinics. These members only have the right to access the blockchain ledger in the global chain by sending a query or access request.


    

  \item \textit{Active Members.} The active members include the supplier, vaccine manufacturer, warehouse, distributor, and government health authorities. These members communicate with each other through their local blockchains and also have the right to access and update the blockchain ledger in the global chain. 

      \item \textit{Decentralised Ledger.} All active members use private blockchain networks. Another blockchain network connects all nodes in the global chain layer. Only authorised and verified supply chain entities in the network can store the vaccine data in the main ledger and create immutable links in the form of blockchain transactions, which are time-stamped and secured using lightweight cryptographic technology. An entity's local blockchain network is where it stores its private and sensitive data. Users no longer need to keep all their data on the global chain, providing an advantage. Consequently, the combination of local and global chains is beneficial, as nodes in the network only retain indexed data, making it easier to locate where the data are stored.

    \item \textit{Smart contracts.} Our proposed framework consists of five smart contracts, namely, the classification contract
    (CC), consortium contract (CTC), validation contract (VLC), the access contract (ACC) and the revoke contract (REC), each of which implements the access control policies for a pair of entities in the vaccine supply chain. 

    \end{itemize}

\subsection{Multi-layer Architecture}

Multi-MedChain uses a multi-layer, multi-blockchain approach to distribute commodities efficiently. It enables atomic data transfer and IP privacy between local blockchain systems. Three layers make up this permissioned blockchain solution. Starting from the bottom up, each layer is added. Each layer is introduced in a bottom-up approach as follows.

\begin{itemize}
    \item \textit{User Layer.} The bottom layer is passive and cannot change the global chain. These members can request vaccine data from the blockchain and report. The patient can request the vaccine production date or other immunization data and report side effects.      Passive users include government hospitals, clinics, and immunization centers that generate and store patients' EMRs in their local blockchain (BC) network. They also communicate with the Government Health Authority-owned local BC. MPC nodes in Multi-MedChain enable secure multi-party communication across layers \cite{goldreich1998secure}. The GHA gives user layer patients a digital COVID-19 vaccine certificate that they can show authorities. The smart contract lets passive user layer members query or access the main global chain.
    
     \item \textit{Local Chain Layer.} Our suggested system's middle layer is a consortium of active members, mostly raw material suppliers, vaccine makers with warehouses, distributors, and GHA. Local BCs of each member securely communicate through local MPC nodes in this tier. These members preserve sensitive IP on their local BCs and exchange just vaccination order placement and shipping tracking data. These members maintain ACLs by mutual agreement. In our suggested method, GHA registers and verifies all entities and sets the ACL for passive user layer members. Due to many entities at this layer, commodities transfers are paired with local or global BC transactions. Each entity can only write certain transactions to the local BC depending on data sensitivity and role in the proposed system. Transactions and data are multi-signed when entities share data with local or global BCs. 
   
    \item \textit{Global Chain Layer.} 
    Topmost layer in our suggested architecture supports user-local layer sharing protocol. Local chain layer members can read and write transactions and share the block header with the global chain layer for verification. Block headers are maintained by global nodes. When a local chain member requests product tracing or tracking information, the global chain is consulted. After verifying the request, the global BC generates a block header owner timestamp. Finally, the requester receives the data.
\end{itemize}
 

\subsection{Transaction Management}

\begin{figure}[tbh]
\centering{}\textsf{\includegraphics[width=1\columnwidth,scale=1]{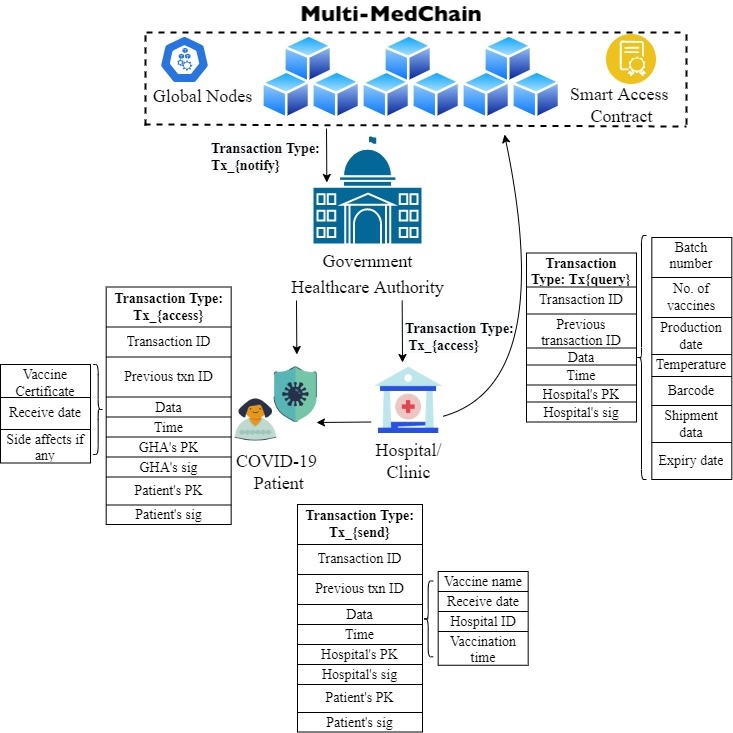}}\caption{Transactional asset flow in user chain layer.} \label{fig: transaction flow in UC}
\end{figure}

\begin{figure}[tbh]
\centering{}\textsf{\includegraphics[width=1\columnwidth,scale=5]{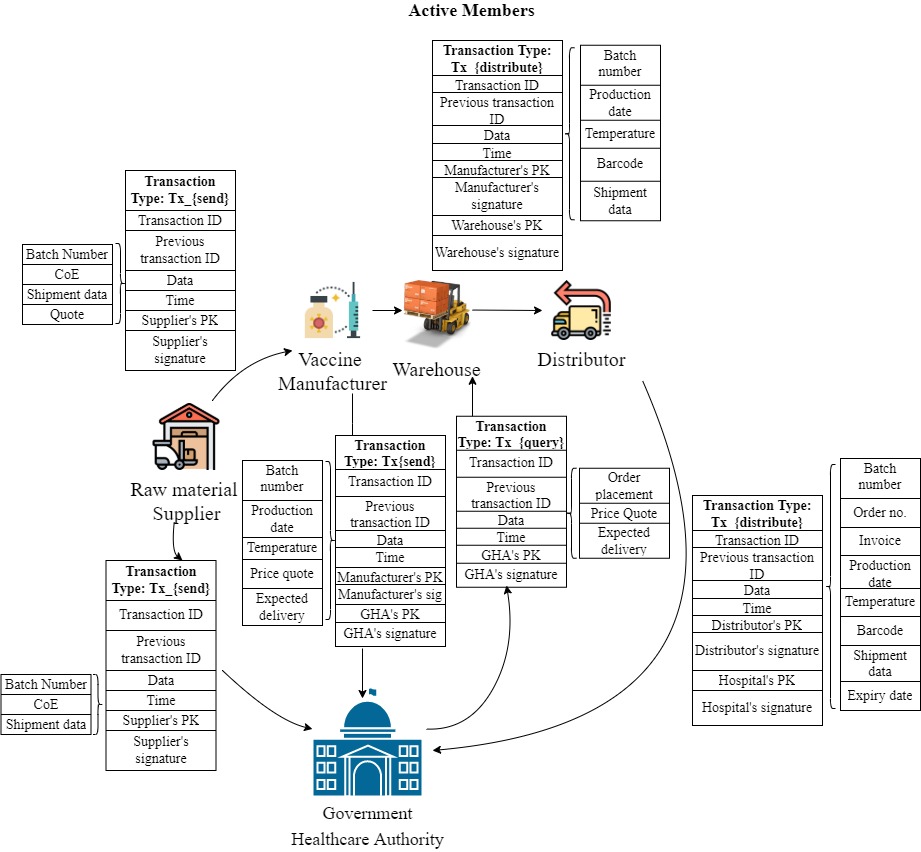}}\caption{Transactional asset flow in local chain layer.} \label{fig: transaction flow in LC}
\end{figure}

The suggested approach focuses on three transaction types: access, send, and query. It helps supply chain entities trade physically. Figs. illustrate that each entity in our architecture can write specific transactions to the user, local, and global chain layers based on its role.See \ref{fig: transaction flow in UC} and \ref{fig: transaction flow in LC}. Users at the passive layer can only write query and access transactions. Transactions in the Vaccine Chain can be single-signed (\textit{unisig}) or multi-signed (\textit{multisig}) based on communication among entities.  It builds consortium trust since transactions require consensus and consent from numerous trusted parties, eliminating conflicts. Multi-signature techniques reduce the danger of single point of failure attacks by dispersing authorization among several parties and preventing malevolent users from taking control. Every vaccination product is stored on the worldwide chain during manufacture, distribution, transfer, and tracking.

\subsection{Smart Contract-based Access Control}

 The proposed method employs smart contracts for access control \cite{sworld}. Members are identified and access rights are defined using ACL \cite{9235494}. It limits communication and access control on incoming searches and access requests to authorized parties (Figure 6). Our fine-grained access control architecture includes five smart contracts: categorization (CC), consortium (CTC), validation (VLC), access (ACC), and revoke (REC). CC lists and classifies vaccination chain entities. Per consortium agreement, CTC records stakeholder agreements and shares access. VLC verifies query requester and block header owner in global BC, but data sender and recipient in local BC. ACC and REC grant and deny data requesters at both tiers. The consortium may apply conditions or agreement policies to these ACLs at any moment.

\begin{algorithm}[tbh]
\begin{description}
\item [{\caption{accountRegister \& Deployment ABI}
}]~
\end{description}
\textbf{Input:} \textit{credential details for registeration}

\textbf{Output}: \textit{contract address, confirmation status
of the registered method.}

Required: {\textit{name, unique email ID, identity of user, password, abi.}}
\begin{enumerate}
\item Select(\textit{entity$\leftarrow$users}) \textbf{then}
\item \textit{Info}$\leftarrow$ \textit{entity}{[}\textit{unique
email ID}{]}
\item \textbf{If} (info == True ) \textbf{then}

\textit{Registration failed (Account already exists)}

\textit{Confirmation}$\leftarrow$ \textbf{false}
\item \textbf{Else}
 
 \textit{Check identity} \textbf{then}

 \textit{Compile corresponding contract} \textbf{then}

 \textit{Deploy contract}

 \textit{Entity}$\leftarrow$\textit{[name], [unique email ID], [identity], [password], [contract address]}

 \textit{Create} (users$\leftarrow$ entity) 

 \textit{Info} $\leftarrow$ contract address

 \textit{Confirmation} $\leftarrow$ \textbf{true} 

\item \textbf{End}
\end{enumerate}
\end{algorithm}

To implement the register functionality, the classification steps are designed and implemented as shown in Algorithm 1. The \textit{delivery()} function stores on-chain shipping information for deliveries, as seen in Algorithm 2.  Delivery from and to addresses are recorded during creation. Then, product amount and kind are included during delivery preparation. Shipping and receiving times are recorded when the delivery is shipped or received. Most applications require manual delivery completion, however our system updates delivery status automatically following inventory changes. The \textit{delivery()} function stores on-chain shipping information for deliveries, as seen in Algorithm 2.  Delivery from and to addresses are recorded during creation. Then, product amount and kind are included during delivery preparation. Shipping and receiving times are recorded when the delivery is shipped or received. Most applications require manual delivery completion, however our system updates delivery status automatically following inventory changes. 

\begin{algorithm}[tbh]
\begin{description}
\item [{\caption{Delivery contract deployment}
}]~
\end{description}
\textbf{Input:} \textit{ address of both receiver and sender, product info}

\textbf{Output}: \textit{contract address, confirmation status.}

Required: {\textit{address\_from, address\_to, product\_id, product\_quantity, producer\_address, time.}}
\begin{enumerate}
\item Compile corresponding contract \textbf{then}
\item \textit{Deploy contract}$\leftarrow$ \textit{address\_from}, \textit{address\_to}

\item \textit{Info}$\leftarrow$ \textit{address\_delivery}

\item \textit{addProduct}$\leftarrow$ \textit{product\_id}, \textit{product\_quantity}, \textit{producer\_address}

\item \textit{shipping}$\leftarrow$ \textit{time}

\item \textit{Entity}$\leftarrow$ [\textit{name}, \textit{address\_delivery}, \textit{address\_from}, \textit{address\_to}, \textit{time}]

 \item \textit{Create} (Delivery$\leftarrow$ Entity) 

 \item \textit{Confirmation} $\leftarrow$ \textbf{true} 

\item \textbf{End}
\end{enumerate}
\end{algorithm}

Algorithm 3 suggests searching the entire supply chain for records to trace the product. To ease this procedure, all deliveries are maintained under product batches, guaranteeing that each batch contains all deliveries and is traceable throughout the supply chain. Tracking gives patients access to full manufacturer-to-hospital delivery information, including departure and arrival times. All records are matched to on-chain data for tamper-proof confirmation, confirming their reliability.

\begin{algorithm}[tbh]
\begin{description}
\item [{\caption{Trace delivery record for patients}
}]~
\end{description}
\textbf{Input:} \textit{product info}

\textbf{Output}: \textit{delivery records}

Required: {\textit{ product name, product production date, product batch number}}
\begin{enumerate}
\item Select(\textit{entity$\leftarrow$items}) \textbf{then}
\item \textit{Info}$\leftarrow$ \textit{entity}{[}\textit{product name, product date, product batch number}{]}
\item \textbf{If} (info == False) \textbf{then}

\textit{Tracking failed (Could not find item)}

\textit{Confirmation}$\leftarrow$ \textbf{false}
\item \textbf{Else}
 
 \item \textit{Delivery detail} $\leftarrow$ Info \textbf{then}

\textit{Select (Entity $\leftarrow$ Delivery)} \textbf{then}

  \item \textit{Entity} $\leftarrow$ [deliveryIds]
  
 \textit{Result} $\leftarrow$ Entity

  \textit{Confirmation} $\leftarrow$ \textbf{true}

\item \textbf{End}
\end{enumerate}
\end{algorithm}

\begin{figure}[tbh]
\vfill
\centering{}\textsf{\includegraphics[width=0.9\columnwidth,scale=1]{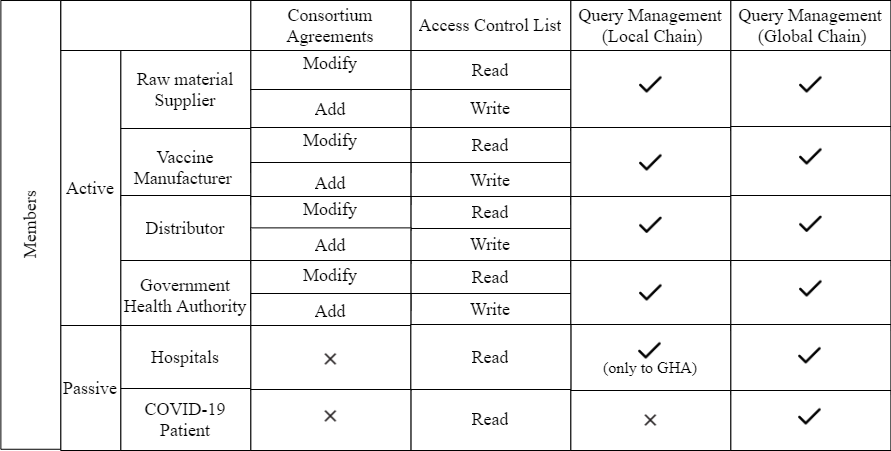}}\caption{Access control list management.} \label{fig: ACL}
\end{figure}

\section{Implementation \& Results}

\subsection{Implementation}
 The implementation is done using the web3 Python library that facilitates interaction with the Ethereum blockchain, allowing us to build decentralised application (DApp) by providing a simple interface to interact with smart contracts and access blockchain data. 

\subsubsection{Ethereum and Ganache}
Ethereum, a decentralised and open source blockchain platform, empowers the creation
and execution of smart contracts and decentralised applications (DApps). 
Our implementation of smart contracts focuses on storing supply chain data on-chain, thereby enabling the functionalities of inventory management and delivery tracking. Ganache, a personal blockchain emulator, establishes a local environment, allowing for the simulation of Ethereum network behaviour on individual machines. 

\subsubsection{React and Database} We use React for front-end development due to its efficiency, flexibility, and developer-friendly features. Additionally, we have employed an SQLite database, where all shipment data has been recorded off-chain.
For the overall view of the system architecture, both the off-chain database and the blockchain would interact with the server. During the data storage and verification process, a comparison between the off-chain and on-chain contents is conducted to ensure that the data are tamper-free.

\subsection{Results}

\begin{figure}[tbh]  
\centering{}\textsf{\includegraphics[width=0.8\columnwidth,scale=1]{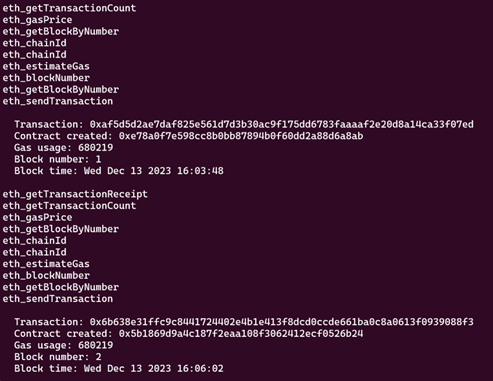}}\caption{Transaction details and gas usage.} \label{fig: CLI}
\end{figure}

Ethereum blockchain platform denotes the amount of work done in the form of a unit called gas. Since we have used the test Ethereum network, these gas values are only test values. In our case study, as depicted in Fig.  7, the gas required to deploy various functions such as \textit{deployment()}, \textit{inbound inventory()}, \textit{send delivery()}, \textit{add product()}, \textit{set shipment()} and \textit{receive()} in regards to manufacturer, wholesaler/hospital and distributor is given in Table II. 

\begin{table}[]
\centering
\caption{Gas deployment cost.}
\resizebox{\columnwidth}{!}{
\begin{tabular}{|l|l|l|l|}
\hline
\textbf{Functions}         & \textbf{Manufacturer} & \textbf{Wholesaler/Hospital} & \textbf{Distributor} \\ \hline
Deployment                 & 680219                & 680219                       & 1029634              \\ \hline
Inbound inventory          & 95956                 & 73411                        & N/A                  \\ \hline
Send Delivery/Outbound     & 1160363               & 1153747                      & N/A                  \\ \hline
Add Product                & N/A                   & N/A                          & 50636                \\ \hline
Set Shipment               & N/A                   & N/A                          & 51413                \\ \hline
Receive                    & N/A                   & N/A                          & 51347                \\ \hline
\end{tabular}
}
\label{Table2}
\end{table}

\subsubsection{Latency Analysis}
Latency heavily depends on the blockchain interaction. For normal requests or functionalities accomplishment, latency generally remains below 20 ms (e.g., Login/Logout function, manufacturer new item/check item function, and tracking functions). Despite the low latency of these simple functionalities, differences arise depending on the number of queries to the local off-chain database. 
However, functionalities that involve interaction with the smart contract experience a dramatic increase in processing time.
The deployment time incurs the most significant cost, leading to considerable latency for the Register function, as shown in Fig. 8(a). Additionally, the inbound functions for manufacturers (Fig. 8(b)) and the outbound functions for hospitals or wholesalers (Fig. 8(c)) are not as large due to a single function call inside the contract. However, latency is high for manufacturer outbound inventory functions and hospital or wholesaler inbound functions due to cross-contract function calls. 


\begin{figure*}[h]
  \centering
  \subfigure[]{\includegraphics[width=2in]{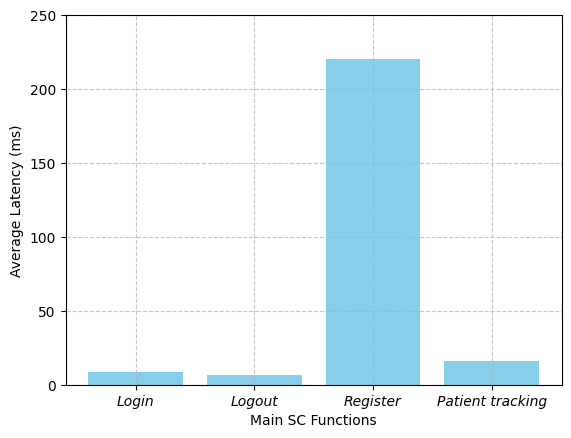}}
  \subfigure[]{\includegraphics[width=2in]{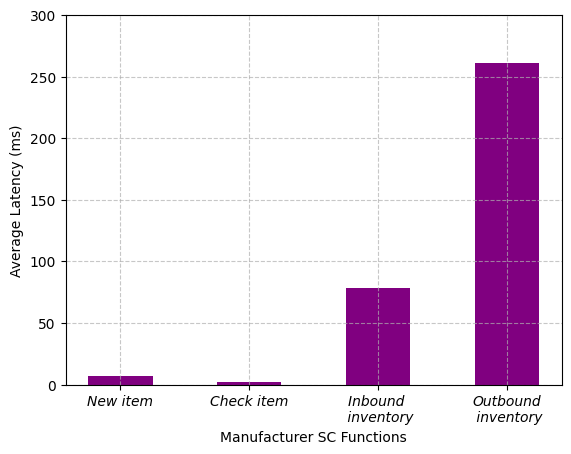}}
   \subfigure[]{\includegraphics[width=2in]{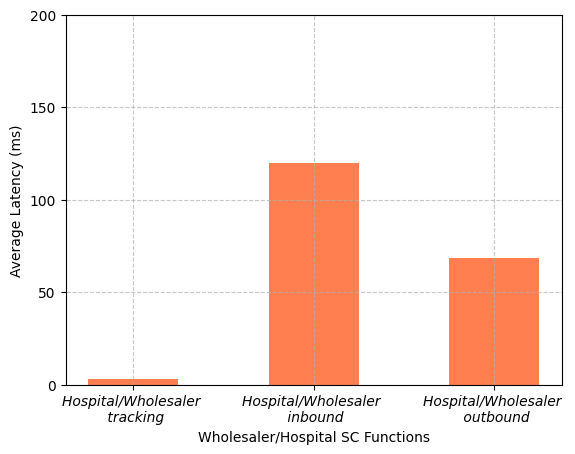}}
  \caption{Latency during the execution of (a) main SC functions, (b) manufacturer SC functions, and (c) wholesaler/hospital SC functions.}
  \label{fig8}
\end{figure*}

Table III compares the planned work to related works [17–20]. Our system is more scalable than \cite{li2023novel} and \cite{peng2020efficient}, mainly storing data off-chain and meeting security requirements. Our system achieves reduced latency for many operations by storing most data off-chain. Compared to \cite{sharma2023blockchain}, our solution reduces gas expenses by optimizing smart contract data format and minimizing on-chain data storage. Mapping is utilized across all smart contracts to optimize data structure and eliminate query loops. With our solution, you can trace the supply chain and manage inventories regardless of the blockchain \cite{musamih2021blockchain}. We maintain reasonable gas costs, matching or surpassing the cost-effectiveness of the proposed system by \cite{musamih2021blockchain} for specific functions.

\begin{table}[]
\centering
\caption{Performance comparison between our proposed solution and related work.}
\resizebox{\columnwidth}{!}{
\begin{tabular}{|l|l|l|l|l|l|}
\hline
\textbf{Feature}            & \textbf{[17]} & \textbf{[18]} & \textbf{[19]} & \textbf{[20]} & \textbf{Proposed system} \\ \hline
Decentralised               & Yes           & Yes           & No            & Yes           & Yes                      \\ \hline
Blockchain                  & Fabric        & Fabric        & Ethereum      & Ethereum      & Ethereum                 \\ \hline
Off-chain storage           & No            & Yes           & No            & Yes           & Yes                      \\ \hline
Latency                     & Medium        & High          & High          & High          & Low                      \\ \hline
Scalability                 & Low           & Low           & Medium        & Medium        & Higher                   \\ \hline
\end{tabular}
}
\label{Table2}
\end{table}

\subsubsection{Features of Proposed System}
The proposed project, in which the government takes ownership of the Multi-MedChain and uses smart contracts for various processes, has several results and business implications.

\begin{itemize}


    \item \textit{Transparency and Accountability:} Supply chain transparency is achieved using blockchain and smart contracts. An immutable ledger records all transactions and interactions, ensuring stakeholder accountability and auditability.

     \item \textit{Enhanced Data Security:} Blockchain's decentralisation and cryptographic security make vaccination order, shipment, and recipient data tamper-resistant and secure.

     \item \textit{Improved Access Control:} Smart contracts regulate who can access data and when in the supply chain. Mutual agreements govern access permissions, improving data privacy and reducing data breaches.

     \item \textit{Faster Order Placement and Fulfillment:} Smart contracts help the government order vaccines faster. Order fulfillment and response time can be accelerated by manufacturers.

     \item \textit{Efficient Allocation:} The distributor's capacity to deploy vaccinations based on GHA guidelines ensures vaccines are supplied according to priority. This ensures high-priority vaccine users receive them quickly.

     \item \textit{Accurate Tracking and Reporting:} Vaccine distribution is tracked accurately via blockchain transactions. Vaccine movement reports and insights help with decision-making and resource allocation.




\end{itemize}






\section{Discussion and Conclusion}



Multi-MedChain facilitates the integration of multiple blockchain platforms within a three-layer multi-party, multi-blockchain framework. It establishes a scalable and distributed SCM system for healthcare, addressing crucial gaps in existing solutions such as end-to-end traceability and collaborative access control to restrict access to private data, which are essential for regulatory and compliance reporting.  Multi-MedChain promotes transparency among stakeholders and facilitates secure communication to ensure trust. We leverage smart contracts to devise an access control mechanism that supports custom-designed access authorizations. This results in verifiable automation and reduced information asymmetries within the supply chain.

Our performance results demonstrate the scalability of the proposed solution. However, incorporating blockchain and decentralised access control still encounters some performance challenges compared to traditional centralised supply chain systems. To minimize latency in processing and retrieving supply chain data, the integration of technologies like edge computing and the Internet of Things may prove beneficial. We plan to extend Multi-MedChain and integrate such technologies into our architecture to further improve the practicality of the proposed solution for real-world adoption.

\section{Source Code \& Implementation}

The GitHub link to the project source code is available  \href{https://github.com/akankshasaini/DMSCMS}{here}.

\section*{Acknowledgement}

The authors acknowledge Prof Babak Abbasi for his thoughtful feedback and suggestions.








\bibliographystyle{IEEEtran}

\bibliography{references}

\begin{thebibliography}{10}
\providecommand{\url}[1]{#1}
\csname url@samestyle\endcsname
\providecommand{\newblock}{\relax}
\providecommand{\bibinfo}[2]{#2}
\providecommand{\BIBentrySTDinterwordspacing}{\spaceskip=0pt\relax}
\providecommand{\BIBentryALTinterwordstretchfactor}{4}
\providecommand{\BIBentryALTinterwordspacing}{\spaceskip=\fontdimen2\font plus
\BIBentryALTinterwordstretchfactor\fontdimen3\font minus
  \fontdimen4\font\relax}
\providecommand{\BIBforeignlanguage}[2]{{%
\expandafter\ifx\csname l@#1\endcsname\relax
\typeout{** WARNING: IEEEtran.bst: No hyphenation pattern has been}%
\typeout{** loaded for the language `#1'. Using the pattern for}%
\typeout{** the default language instead.}%
\else
\language=\csname l@#1\endcsname
\fi
#2}}
\providecommand{\BIBdecl}{\relax}
\BIBdecl

\bibitem{park2020global}
C.-Y. Park, K.~Kim, and S.~Roth, \emph{Global shortage of personal protective
  equipment amid COVID-19: supply chains, bottlenecks, and policy
  implications}.\hskip 1em plus 0.5em minus 0.4em\relax Asian Development Bank,
  2020, no. 130.

\bibitem{bray2012information}
R.~L. Bray and H.~Mendelson, ``Information transmission and the bullwhip
  effect: An empirical investigation,'' \emph{Management Science}, vol.~58,
  no.~5, pp. 860--875, 2012.

\bibitem{nakasumi2017information}
M.~Nakasumi, ``Information sharing for supply chain management based on block
  chain technology,'' in \emph{2017 IEEE 19th conference on business
  informatics (CBI)}, vol.~1.\hskip 1em plus 0.5em minus 0.4em\relax IEEE,
  2017, pp. 140--149.

\bibitem{dutta2020blockchain}
P.~Dutta, T.-M. Choi, S.~Somani, and R.~Butala, ``Blockchain technology in
  supply chain operations: Applications, challenges and research
  opportunities,'' \emph{Transportation research part e: Logistics and
  transportation review}, vol. 142, p. 102067, 2020.

\bibitem{pournader2020blockchain}
M.~Pournader, Y.~Shi, S.~Seuring, and S.~L. Koh, ``Blockchain applications in
  supply chains, transport and logistics: a systematic review of the
  literature,'' \emph{International Journal of Production Research}, vol.~58,
  no.~7, pp. 2063--2081, 2020.

\bibitem{kshetri20181}
N.~Kshetri, ``1 blockchain’s roles in meeting key supply chain management
  objectives,'' \emph{International Journal of Information Management},
  vol.~39, pp. 80--89, 2018.

\bibitem{cworld}
``Modum: A seamless and trusted block data exchange,''
  \url{https://www.modum.io/}, [Online].

\bibitem{dworld}
``Chronicled,'' \url{https://www.chronicled.com/}, [Online].

\bibitem{eworld}
``Blockpharma,'' \url{https://www.blockpharma.com/}, [Online].

\bibitem{reda2020blockchain}
M.~Reda, D.~B. Kanga, T.~Fatima, and M.~Azouazi, ``Blockchain in health supply
  chain management: State of art challenges and opportunities,'' \emph{Procedia
  Computer Science}, vol. 175, pp. 706--709, 2020.

\bibitem{jamil2019novel}
F.~Jamil, L.~Hang, K.~Kim, and D.~Kim, ``A novel medical blockchain model for
  drug supply chain integrity management in a smart hospital,''
  \emph{Electronics}, vol.~8, no.~5, p. 505, 2019.

\bibitem{yong2020intelligent}
B.~Yong, J.~Shen, X.~Liu, F.~Li, H.~Chen, and Q.~Zhou, ``An intelligent
  blockchain-based system for safe vaccine supply and supervision,''
  \emph{International Journal of Information Management}, vol.~52, p. 102024,
  2020.

\bibitem{alam2021challenges}
S.~T. Alam, S.~Ahmed, S.~M. Ali, S.~Sarker, G.~Kabir \emph{et~al.},
  ``Challenges to covid-19 vaccine supply chain: Implications for sustainable
  development goals,'' \emph{International Journal of Production Economics},
  vol. 239, p. 108193, 2021.

\bibitem{li2023novel}
J.~Li, D.~Han, Z.~Wu, J.~Wang, K.-C. Li, and A.~Castiglione, ``A novel system
  for medical equipment supply chain traceability based on alliance chain and
  attribute and role access control,'' \emph{Future Generation Computer
  Systems}, vol. 142, pp. 195--211, 2023.

\bibitem{peng2020efficient}
S.~Peng, X.~Hu, J.~Zhang, X.~Xie, C.~Long, Z.~Tian, and H.~Jiang, ``An
  efficient double-layer blockchain method for vaccine production
  supervision,'' \emph{IEEE transactions on nanobioscience}, vol.~19, no.~3,
  pp. 579--587, 2020.

\bibitem{musamih2021blockchain}
A.~Musamih, K.~Salah, R.~Jayaraman, J.~Arshad, M.~Debe, Y.~Al-Hammadi, and
  S.~Ellahham, ``A blockchain-based approach for drug traceability in
  healthcare supply chain,'' \emph{IEEE access}, vol.~9, pp. 9728--9743, 2021.

\bibitem{sharma2023blockchain}
P.~Sharma, S.~Namasudra, N.~Chilamkurti, B.-G. Kim, and R.~Gonzalez~Crespo,
  ``Blockchain-based privacy preservation for iot-enabled healthcare system,''
  \emph{ACM Transactions on Sensor Networks}, vol.~19, no.~3, pp. 1--17, 2023.

\bibitem{yang2021behavioural}
Y.~Yang, J.~Lin, G.~Liu, and L.~Zhou, ``The behavioural causes of bullwhip
  effect in supply chains: A systematic literature review,''
  \emph{International Journal of Production Economics}, vol. 236, p. 108120,
  2021.

\bibitem{xu2019systematic}
M.~Xu, X.~Chen, and G.~Kou, ``A systematic review of blockchain,''
  \emph{Financial Innovation}, vol.~5, no.~1, pp. 1--14, 2019.

\bibitem{monrat2019survey}
A.~A. Monrat, O.~Schel{\'e}n, and K.~Andersson, ``A survey of blockchain from
  the perspectives of applications, challenges, and opportunities,'' \emph{IEEE
  Access}, vol.~7, pp. 117\,134--117\,151, 2019.

\bibitem{gong2021multiple}
Y.~Gong, Y.~Jiang, and F.~Jia, ``Multiple multi-tier sustainable supply chain
  management: a social system theory perspective,'' \emph{International Journal
  of Production Research}, pp. 1--18, 2021.

\bibitem{collier2021zero}
Z.~A. Collier and J.~Sarkis, ``The zero trust supply chain: Managing supply
  chain risk in the absence of trust,'' \emph{International Journal of
  Production Research}, pp. 1--16, 2021.

\bibitem{xu2020data}
S.~Xu and K.~H. Tan, ``Data-driven inventory management in the healthcare
  supply chain,'' in \emph{Supply Chain and Logistics Management: Concepts,
  Methodologies, Tools, and Applications}.\hskip 1em plus 0.5em minus
  0.4em\relax IGI Global, 2020, pp. 1390--1403.

\bibitem{clauson2018leveraging}
K.~A. Clauson, E.~A. Breeden, C.~Davidson, and T.~K. Mackey, ``Leveraging
  blockchain technology to enhance supply chain management in healthcare:: An
  exploration of challenges and opportunities in the health supply chain,''
  \emph{Blockchain in healthcare today}, 2018.

\bibitem{goldreich1998secure}
O.~Goldreich, ``Secure multi-party computation,'' \emph{Manuscript. Preliminary
  version}, vol.~78, no. 110, 1998.

\bibitem{sworld}
``Introduction to smart contracts,''
  \url{https://docs.soliditylang.org/en/v0.8.6/introduction-to-smart-contracts.html},
  [Online].

\bibitem{9235494}
A.~Saini, Q.~Zhu, N.~Singh, Y.~Xiang, L.~Gao, and Y.~Zhang, ``A
  smart-contract-based access control framework for cloud smart healthcare
  system,'' \emph{IEEE Internet of Things Journal}, vol.~8, no.~7, pp.
  5914--5925, 2021.

\end{thebibliography}
\end{document}